# Electrically Tunable Wafer-Sized Three-Dimensional Topological Insulator Thin Films Grown by Magnetron Sputtering

Qixun Guo[1], Yu Wu[1], Longxiang Xu[1], Yan Gong[2], Yunbo Ou[2], Yang Liu[1], Leilei Li[1], Jiao Teng[1]*, Yu Yan[3]*, Gang Han[4], Dongwei Wang[5], Lihua Wang[6], Shibing Long[7], Bowei Zhang[8], Xun Cao[8], Shanwu Yang[4], Xuemin Wang[4], Yizhong Huang[8], Tao Liu[9], Guanghua Yu[1], Ke He[2]*

[1]Department of Material Physics and Chemistry, University of Science and Technology Beijing, Beijing 100083, P. R. China

[2]State Key Laboratory of Low-Dimensional Quantum Physics, Tsinghua University, Beijing 100084, P. R. China

[3]Corrosion and Protection Center, Key Laboratory for Environmental Fracture (MOE), Institute of Advanced Materials and Technology, University of Science and Technology Beijing, Beijing 100083, P. R. China

[4]Collaborative Innovation Center of Advanced Steel Technology, University of Science and Technology Beijing, Beijing 100083, P. R. China

[5]CAS Key Laboratory of Standardization and Measurement for Nanotechnology, National Center for Nanoscience and Technology, Beijing 100190, P. R. China

[6]Institute of Microstructure and Property of Advanced Materials, Beijing Key Lab of Microstructure and Property of Advanced Materials, Beijing University of Technology, Beijing, 100124, P. R. China

[7]School of Microelectronics, University of Science and Technology of China, Hefei, 230026, P. R. China

[8]School of Materials Science and Engineering, Nanyang Technological University, 50 Nanyang Avenue, 639798, Singapore

[9]Department of Physics, The Ohio State University, Columbus, Ohio 43210, USA

Abstract: Three-dimensional (3D) topological insulators (TIs) are candidate materials for various electronic and spintronic devices due to their strong spin-orbit coupling and unique surface electronic structure. Rapid, low-cost preparation of large-area TI thin films compatible with conventional semiconductor technology is key to the practical applications of TIs. Here, we show that wafer-sized $Bi_2Te_3$ family TI and magnetic TI films with decent quality and well-controlled composition and properties can be prepared on amorphous $SiO_2/Si$ substrates by magnetron cosputtering. The $SiO_2/Si$ substrates enable us to electrically tune $(Bi_{1-x}Sb_x)_2Te_3$ and Cr-doped $(Bi_{1-x}Sb_x)_2Te_3$ TI films between p-type and n-type behavior and thus study the phenomena associated with topological surface states, such as the quantum anomalous Hall effect (QAHE). This work significantly facilitates the fabrication of TI-based devices for electronic and spintronic applications.

Three-dimensional (3D) topological insulators (TIs) have a bulk gap and gapless surface states including an odd number of Dirac cones in a surface Brillouin zone.[1,2] The topological surface states are spin-momentum-locked and are protected by time-reversal symmetry from perturbations such as structural defects, disorder and nonmagnetic impurities. Various exotic quantum effects have been observed or predicted in TI-based materials or structures, such as the quantum anomalous Hall effect (QAHE), topological magnetoelectric effect and chiral Majorana superconductivity. These effects can be used to develop low-energy-consumption electronic devices and topological quantum computers. Recently, TIs have also attracted much attention for their possible applications in spintronic devices. [3-12]

Integrating TIs into the mature semiconductor technology is of key importance to realize their full potential for electronic or spintronic applications. Currently, 3D TI materials are mainly prepared by bulk Bridgman growth[13-15], chemical vapor deposition (CVD)[16,17] or molecular beam epitaxy (MBE)[18-20]. It is difficult to obtain large-area 3D TI films with well-controlled properties by Bridgman growth, and CVD-grown TI films usually need to be transferred from the growth substrate onto other substrates for various electronic devices. On the other hand, the low growth rate and high cost (especially for large wafers) of MBE make it less favored for the mass production of TI-based materials and devices. Magnetron sputtering is a low-cost, high-yield growth method compatible with conventional semiconductor technology. The method is particularly capable of fabricating thin films with complex structures and compositions, which is crucial for applications of TI materials in various devices. However, samples grown by magnetron sputtering are polycrystalline and usually have rather low carrier mobility, which restricts the applications of this technique to the growth of semiconductor materials. [21, 22]

In this study, we grew wafer-sized $Bi_2Te_3$ family TI films on usual amorphous $SiO_2$/Si substrates through magnetron sputtering and found that due to the layered structure of the materials, the films have decent quality with a carrier mobility up to 310 cm$^2$/Vs, comparable to that of the films grown with MBE. Topological surface states were observed in the samples with angle-resolved photoemission spectroscopy (ARPES). The $SiO_2$/Si substrates enable us to gate-tune the carrier density of TI films grown on them, and in $(Bi,Sb)_2Te_3$ films with an appropriate Bi/Sb ratio, the ambipolar field effect is realized. With magnetron sputtering, we also realized the growth of magnetically doped TI $Cr_y(Bi_xSb_{1-x})_{2-y}Te_3$ (CBST) films that show ferromagnetism with an out-of-plane easy magnetization axis. An anomalous Hall resistance ($R_{AH}$) up to 16.4 kΩ was observed at 2 K, promising the observation of the QAHE in the films. These results demonstrate magnetron sputtering as a suitable mass production method for growing TI-based films for various electronic and spintronic applications.

We grew $(Bi_{1-x}Sb_x)_2Te_3$ films by cosputtering $Bi_2Te_3$ and $Sb_2Te_3$ alloy targets as well as Te targets on amorphous $SiO_2$/Si(100) substrates, which were kept at 160 ℃ during growth. By modifying the Bi/Sb ratio in the films, one can control the density as well as the type of carriers in the films. The Bi/Sb ratio in the films was controlled by the sputtering powers of the $Bi_2Te_3$ and $Sb_2Te_3$ target guns. The Te target was also turned on at the same time to reduce the Te deficiency in the films. After growth, the films were annealed at 160 ℃ for 25 minutes. **Figure 1**a shows a photograph of a 10 cm (4 inch) diameter silicon wafer with a 10 nm thick $Bi_2Te_3$ film grown on it, which maintains uniform properties across the whole film. Figure 1b displays a high-resolution transmission electron microscopy (HRTEM) cross-sectional image of a $(Bi_{1-x}Sb_x)_2Te_3$ film, which shows the characteristic quintuple-layer structure of $Bi_2Te_3$. Although the film is polycrystalline, all the grains have the same *c* orientation (normal to the cleavage plane). The crystalline structure of the BST film was further confirmed by X-ray diffraction (XRD) measurements (Figure 1d). The presence of only (0, 0, 3n) diffraction peaks suggests that the film plane is parallel to the cleavage plane. This postulation is reasonable because the material has a layered structure, which means that the cleavage surface has a quite low surface free energy. The above structural characterization results demonstrate that we can indeed obtain $Bi_2Te_3$ family TI films with decent crystalline quality via magnetron sputtering.

We used ARPES to assess the electronic band structure of a 10 QL-thick $Bi_2Te_3$ film grown by magnetron sputtering. Although the film is composed of domains of various in-plane crystalline orientations, the different domains have similar surface state band dispersion near the Dirac point because of the isotropy of the topological s surface states around here, which will give observable ARPES signals. We indeed identified energy bands with similar band dispersion to the topological surface states of $Bi_2Te_3$ (Figure 1b). The observation further confirms that we obtained $Bi_2Te_3$ TI films with magnetron sputtering, although the randomly oriented grains significantly broaden the spectra.

For transport studies, we prepared three $(Bi_{1-x}Sb_x)_2Te_3$ samples, BST1, BST2 and BST3, with x = 0.74, 0.72 and 0.58, respectively. **Figure 2**b shows the $R_{xx}$-*T* curve of the BST2 film. At high temperatures (in the region of ~88-300 K), $R_{xx}$ increases as T is decreased, showing semiconductor-like behavior and indicating that $E_F$ is in the bulk band gap. The $R_{xx}$-*T* curve displays metallic behavior in the intermediate temperature region (~7-88 K), which can be the result of reduced electron-phonon scattering of the surface states.[15, 23, 24] A second increase in $R_{xx}$ when *T* is lower than 7 K can be ascribed to the quantum correlations to conduction, [15], which is similar to those of the MBE-grown films of similar composition and thickness.[23, 24]

Moreover, Figure 2c shows the Hall resistance of sample BST2 as a function of magnetic field $R_{xy}(\mu_0H)$ at various gate voltages. The $R_{xy}$-$\mu_0H$ curves are always linear over the entire range of the magnetic field (±2 T), and the positive slopes of $R_{xy}$-$\mu_0H$ as the gate voltage $V_g$ increases from -120 V to -15 V suggest that the dominant carriers are p-type, while the negative slopes of $R_{xy}$-$\mu_0H$ as $V_g$ continues to increase show that the dominant carriers are n-type. The zero-field longitudinal resistance $R_{xx}$ and Hall coefficient $R_H$ extracted from $R_{xy}$-$\mu_0H$ curves of sample BST1-3 are plotted in Figure 2d-f as a function of gate voltage. In the case of BST1 (BST3), the $R_{xx}$ monotonically increases (decreases) with increasing $V_g$, and the $R_H$ has positive (negative) signs, indicating that BST1 (BST3) is a p-type (n-type) semiconductor. On the other hand, in sample BST2, when $V_g$ is approximately 0 V, $R_H$ reverses its sign, while $R_{xx}$ reaches its maximum value. These results clearly demonstrate the ambipolar field effect of Dirac dispersions and suggest that the Fermi level can be controlled across the DP by tuning the ratio of Bi to Sb (Figure 2a) and back-gate voltages. We note that for sample BST2, the maximum value of $R_{xx}$, corresponding to the gate voltage at the charge neutral point, is on the same order as the maximum values of BST samples deposited by CVD or MBE, as listed in table 1, and is also in the same order as the quantum resistance. The 2D carrier densities $n_{2D}=1/(eR_H)$ (where $e$ is the elementary charge) can be obtained from the linear ordinary Hall resistance, and the lowest carrier density is $n_{2D} = 2.7 \times 10^{12}$ cm$^{-2}$ at $V_g$=60 V with a corresponding mobility of 270 cm$^2$/Vs for electrons, while the lowest carrier density is $n_{2D} = 2.3 \times 10^{12}$ cm$^{-2}$ at $V_g$= -120 V with a corresponding mobility of 310 cm$^2$/Vs for holes. It can be concluded that magnetron sputtering is an effective method to prepare BST thin films with insulating bulk and low carrier densities, while the highly tunable Fermi level on SiO$_2$ allows us to explore the novel properties of topological surface states near the DP.

After the BST thin films with bulk insulating properties were successfully prepared, we focused on the Cr-doped TI thin films grown by magnetron sputtering to investigate the possibility of realizing long-range ferromagnetic order and tuning $E_F$ inside the surface gap. One way to generate long-range ferromagnetic order in TIs is magnetic doping, and magnetically doped TI materials have been grown by CVD,[25] bulk Bridgman growth[26-29] and MBE[30]. Until now, however, magnetron sputtering has rarely been adopted for this purpose. In our experiment, CBST films were prepared by the cosputtering of $Sb_2Te_3$, $Bi_2Te_3$, Te and Cr targets at a substrate temperature of 175 ℃. **Figure 3** presents an example of ferromagnetic properties observed in a 6 nm thick CBST film. At $T$=2 K, the conventional butterfly-shaped $R_{xx}$-$\mu_0H$ curve and nearly square-shaped Hall hysteresis loop with a coercive field of 340 Oe shown in Figure 3a–b suggest that the long-range ferromagnetic order with perpendicular magnetic anisotropy is developed at low temperatures. A Curie temperature ($T_c$) of 11 K is determined by an Arrott plot (Figure 3c), in which $R_{xy}^2$ is plotted against $\mu_0H/R_{xy}$ and the extrapolated intercept is proportional to the saturation magnetization.

Generally, the Hall resistance of the anomalous Hall effect is expressed as:[30, 31]

$$R_{xy} = R_0H + R_{AH}(M) \qquad (1)$$

where $R_0$ is the slope of the ordinary Hall background, $H$ is the applied magnetic field, and $M$ is the magnetization component in the perpendicular direction. **Figure 4**a and b present the longitudinal sheet resistance $R_{xx}$ and Hall resistance $R_{xy}$, respectively, as a function of $H$ of a CBST thin film at various back-gate voltages. $R_{AH}$ estimated by the intercept of the linear background at a high magnetic field and zero field $R_{xx}(0)$ are plotted in Figure 4c. We observe that the $R_{AH}$ and $R_{xx}(0)$ data point towards the same trend and that both reach their maximum at $V_g$=-130 V and their minimum at $V_g$=150 V. It is worth noting that the maximum of $R_{AH}$ in Figure 4c is 16.4 kΩ, which exceeds 60% of the quantum Hall resistance ($h/e^2$). More importantly, the existence of robust ferromagnetic order in both p-type and n-type Cr-doped TI thin films grown by magnetron sputtering and the $V_g$-independent coercive field shown in Figure 4b demonstrate the occurrence of carrier-independent bulk van Vleck magnetism, which is consistent with the reports in the literature.[30, 32]

Nevertheless, the QAHE has not been observed in our Cr-doped TI films grown by magnetron sputtering. In the quantum anomalous Hall regime, $R_{AH}$ should reach a maximum, and $R_{xx}$ should exhibit a dip at the charge neutral point.[20, 33, 34] This signature has not been observed in our CBST thin films at $T$=2 K. The $R_{AH}/R_{xx}(0)$ ratio of the CBST thin film, however, has a maximum of approximately 0.2 at $V_g$=20 V (see Supplementary Information S6), which exceeds the Hall angle of usual diluted magnetic semiconductors.[30] Therefore, the quantum anomalous edge states in CBST might contribute to carrier transport. We believe that it is possible to observe the QAHE in CBST thin films grown by magnetron sputtering if the quality of the sample is improved further by reducing the sputtering rate, controlling the amount of Cr more precisely and adjusting the annealing process, among other measures.

In summary, we have demonstrated that it is possible to grow $Bi_2Te_3$ family TI thin films on amorphous $SiO_2$/Si substrates by magnetron sputtering. Our work provides a large-scale method to produce bulk insulating TI thin films with tunable transport properties. Both the ARPES data and the ambipolar field effect point to the conclusion that the topological surface states can very well exist in these sputtered, polycrystalline thin films. An unusually large value of 60% of the quantum anomalous Hall resistance is observed in Cr-doped BST films, and more efforts are needed to obtain an ideal quantum anomalous Hall insulator. The magnetron cosputtering growth provides a first yet important step towards the large-scale fabrication of multilayer films with complex structures for future spintronic devices.

**Methods**

Magnetron sputtering growth. $Bi_2Te_3$, BST and CBST thin films were grown on polished, thermally oxidized silicon substrates (300 nm $SiO_2$/silicon) by magnetron sputtering. The cosputtering method with a high-purity (99.99%) $Bi_2Te_3$ alloying target, $Sb_2Te_3$ alloying target and Te target was adopted to control the films' composition. The BST thin films were deposited from $Bi_2Te_3$ and $Sb_2Te_3$ targets by direct current (DC) sputtering and from a Te target by radio-frequency (RF) sputtering. The CBST thin films were deposited from $Bi_2Te_3$, $Sb_2Te_3$ and Cr (99.95%) targets by DC sputtering and a Te target by RF sputtering. During growth, the Si(100) substrate was kept at 160 ℃ for BST thin films and at 175 ℃ for CBST thin films. The base pressure of the deposition chamber was below $3 \times 10^{-7}$ Torr, and the working argon pressure was set at $2 \times 10^{-3}$ Torr. After growth, the thin films were annealed for 25 minutes at the same temperature as the deposition. When the substrate temperature was cooled to room temperature, a 3-4 nm thick Al thin film was sputtered *in situ* on the topological insulator thin films, and the Al thin film was naturally oxidized to form $Al_2O_3$ after the sample was removed from the chamber and exposed in air.

Material characterization. To avoid possible contamination of $Bi_2Te_3$ thin films, a 2 nm thick Te capping layer was deposited on top of the films before we removed them from the high-vacuum growth chamber. The $Bi_2Te_3$ thin films with a 2 nm Te capping layer were exposed in air for 2 hours before ARPES measurements. ARPES data were collected by a Scienta R4000 analyzer at a 3 K sample temperature. The structural characteristics of BST films were investigated by using TEM (FEI Tecnai F20; acceleration voltage, 200 kV) and XRD (Rigaku TTR-Ⅲ, Japan).

Transport measurement. The $Bi_2Te_3$, BST and CBST thin films were mechanically scratched into six-terminal Hall bar configurations by a sharp metal tip; then, aluminum wires were glued to each terminal with silver epoxy to form ohmic contacts. The $R_{xx}$ and $R_{xy}$ data were symmetrized and antisymmetrized, respectively, to account for the misalignment of electrodes. The transport measurements were carried out using a 9 T Physical Properties Measurement System (PPMS, Quantum Design) with a base temperature of 2 K.

Acknowledgements

This work was supported by the National key R & D plan program of China (Grant No. 2017YFF0206104, 2014GB120000), the National Key Scientific Research Projects of China (Grants No.2015CB921502), the Natural Science Foundation of China (Grant Nos. 61574169, 61474007, 51331002), Beijing Laboratory of Metallic Materials and Processing for Modern Transportation, the Opening Project of Key Laboratory of Microelectronics Devices & Integrated Technology, Institute of Microelectronics of Chinese Academy of Sciences.

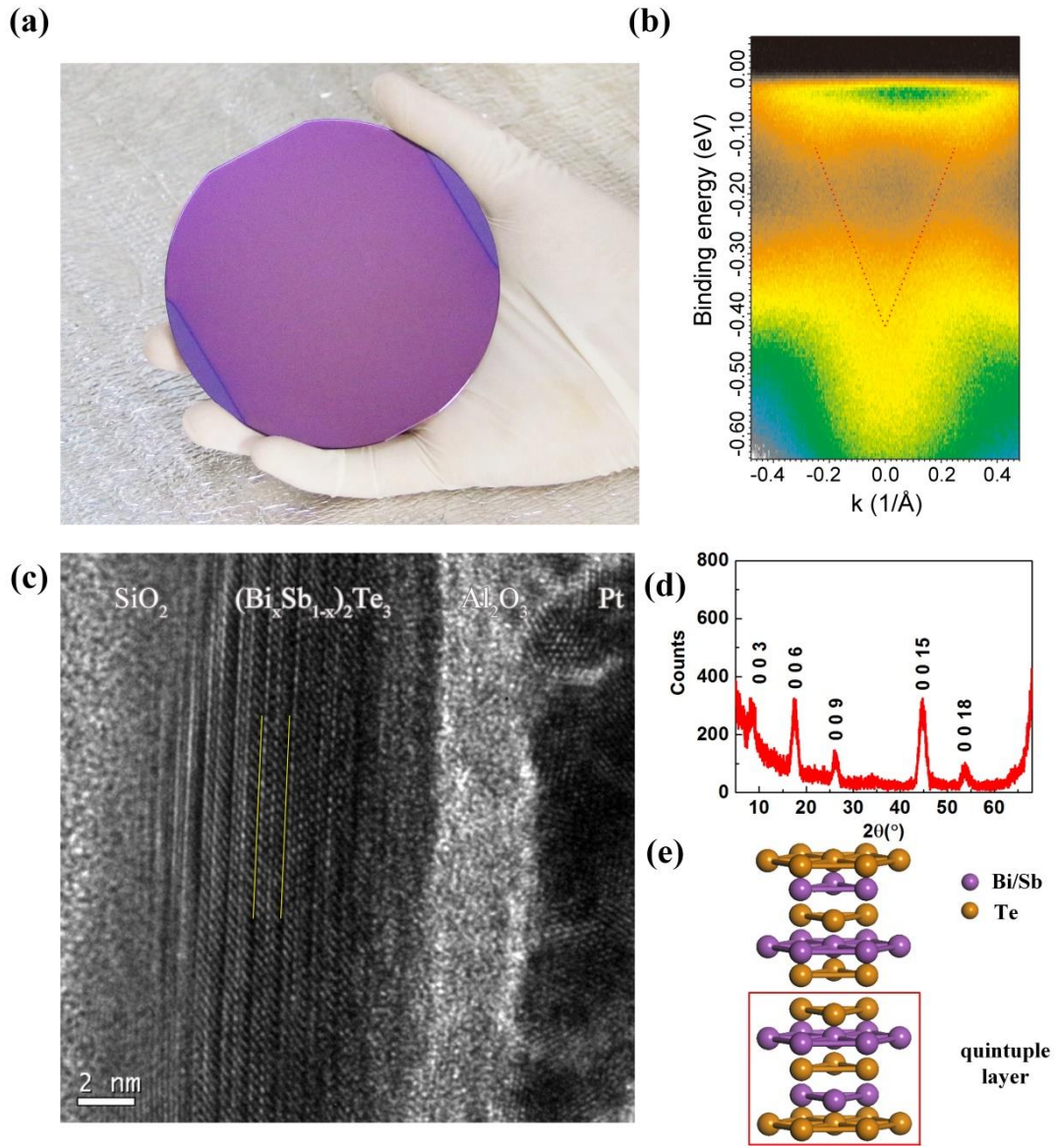

**Figure 1.** Structural characterization of $Bi_2Te_3$ and BST thin films grown on amorphous $SiO_2$/Si by magnetron sputtering. a) Photograph of a $Bi_2Te_3$/$SiO_2$ film with a diameter of 10 cm. b) Electronic band structure of a $Bi_2Te_3$ thin film measured by ARPES showing linear dispersion of topological surface states (red dashed line). c) HRTEM image and d) XRD pattern of a BST thin film grown on amorphous $SiO_2$ (sample BST2). e) The rhombohedral crystal structure of $Bi_2Te_3$ and $(Bi_{1-x}Sb_x)_2Te_3$.

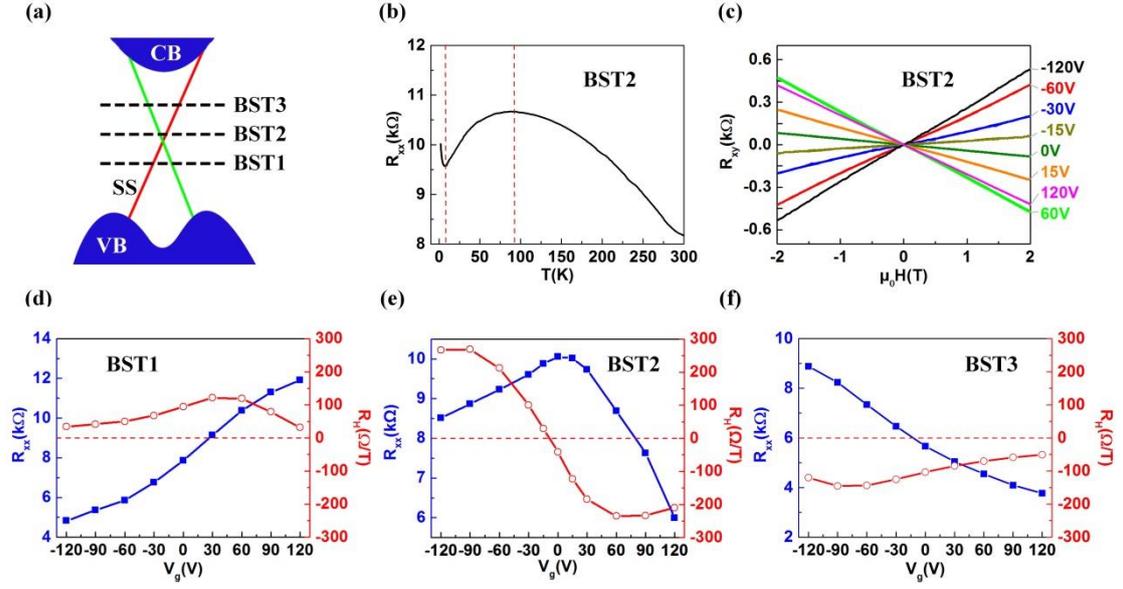

**Figure 2.** Electrical transport in 7 nm thick BST thin films. a) Schematic band structure for samples BST1, BST2 and BST3 and the corresponding Fermi levels, where CB, VB and SS are the bulk conduction band, bulk valance band and surface state, respectively. b) The temperature (T) dependence of the longitudinal sheet resistance ($R_{xx}$) at zero magnetic field. From left to right, two red dashed lines are located at $T$=7 K and $T$=88 K, respectively. c) Hall resistance $R_{xy}$ as a function of magnetic field $\mu_0 H$ at 2 K tuned by the back-gate voltage. d-f) Dependence of zero field longitudinal sheet resistance $R_{xx}(0)$ (solid squares) and Hall coefficient $R_H$ (open circles) on the back-gate voltage for samples BST1(d), BST2(e) and BST3(f) at $T$=2 K.

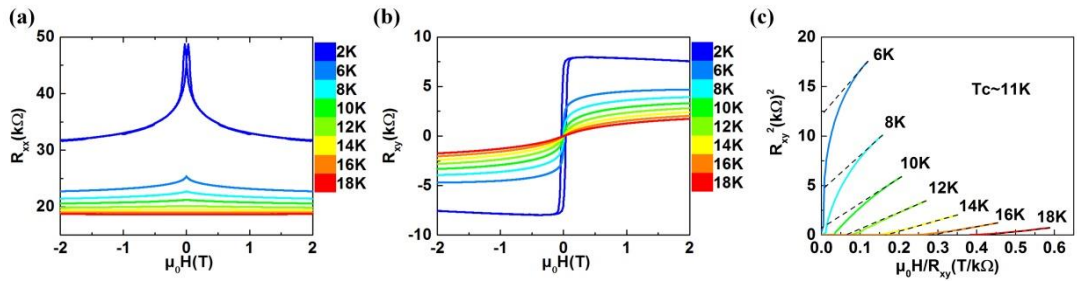

**Figure 3.** Ferromagnetic characterization of a 6 nm thick CBST thin film. a) Longitudinal resistance $R_{xx}$ and b) Hall resistance $R_{xy}$ vs $\mu_0 H$ at different temperatures. c) Arrott plot of the CBST thin film.

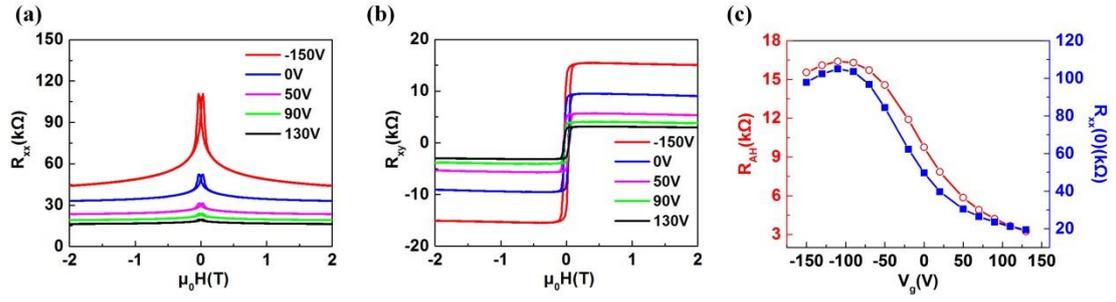

**Figure 4.** a) Longitudinal sheet resistance $R_{xx}$ and b) Hall resistance $R_{xy}$ vs $\mu_0 H$ at different gate voltages at a temperature of 2 K. c) $V_g$-dependent $R_{AH}$ (open circles) and $R_{xx}(0)$ (solid squares) measured at 2 K.

| Table 1 Gate-tuned longitudinal sheet resistance of BST samples developed by different techniques. | | | | |
| --- | --- | --- | --- | --- |
| Device | $R_{xx}$ (kΩ) | T (K) | Thickness (nm) | Reference |
| CVD-grown BST | 0.5-5.5 | 2 | 5 | [35] |
| MBE-grown BST | 8-12 | 2.2 | 20 | [36] |
|  | 3-12 | 0.04 | 8 | [37] |
|  | 2-6 | 2.1 | 7 | [38] |
|  | 2-8 | 1.6 | 8 | [38] |
|  | 1-3 | 1.6 | 20 | [23] |
|  | 2-12 | 1.4 | 10 | [39] |
|  | 3-12 | 6.9 | 10 | [40] |
| Sputter-deposited BST | 6-10 | 2 | 7 | This work |